\documentclass[journal=ancac3,manuscript=article]{achemso}
\usepackage{graphicx}
\usepackage[english]{babel} 
\usepackage[hidelinks, pdftex]{hyperref}
\usepackage{subfigure}
\usepackage{xcolor}
\usepackage{pdfpages}
\DeclareUnicodeCharacter{2212}{-}

\author{Marcel S. Claro}
\email{marcel.santos@usc.es}
\affiliation{CiQUS, Centro Singular de Investigacion en Quimica Bioloxica e Materiais Moleculares, Departamento de Quimica-Fisica, Universidade de Santiago de Compostela, 15782-Santiago de Compostela, Spain.}
\author{Javier Corral-Sertal}
\affiliation{CiQUS, Centro Singular de Investigacion en Quimica Bioloxica e Materiais Moleculares, Departamento de Quimica-Fisica, Universidade de Santiago de Compostela, 15782-Santiago de Compostela, Spain.}
\author{Adolfo O. Fumega}
\affiliation{Department of Applied Physics, Aalto University, FI-00076 Aalto, Finland}
\author{Santiago Blanco-Canosa}
\affiliation{Donostia International Physics Center (DIPC), 20018 San Sebastián, Spain}
\alsoaffiliation{IKERBASQUE, Basque Foundation for Science, Bilbao, Spain}
\author{Manuel Suárez-Rodríguez}
\affiliation{CIC nanoGUNE BRTA, Donostia-San Sebastián, Spain}
   \author{Luis E.  Hueso}
   \affiliation{CIC nanoGUNE BRTA, Donostia-San Sebastián, Spain}
\alsoaffiliation{IKERBASQUE, Basque Foundation for Science, Bilbao, Spain}
\author{Victor Pardo}
\email{victor.pardo@usc.es}
\affiliation{Departamento de Física Aplicada, Universidade de Santiago de Compostela, E-15782 Santiago de Compostela, Spain}
\alsoaffiliation{Instituto de Materiais iMATUS,
  Universidade de Santiago de Compostela, E-15782 Campus Sur s/n,
  Santiago de Compostela, Spain}
\author{Francisco Rivadulla}
\affiliation{CiQUS, Centro Singular de Investigacion en Quimica Bioloxica e Materiais Moleculares, Departamento de Quimica-Fisica, Universidade de Santiago de Compostela, 15782-Santiago de Compostela, Spain.}

\title{Temperature and thickness dependence of the thermal conductivity in 2D ferromagnet Fe$_3$GeTe$_2$}

\AtEndDocument{\includepdf[pages={-},pagecommand={\thispagestyle{headings}},scale=0.9]{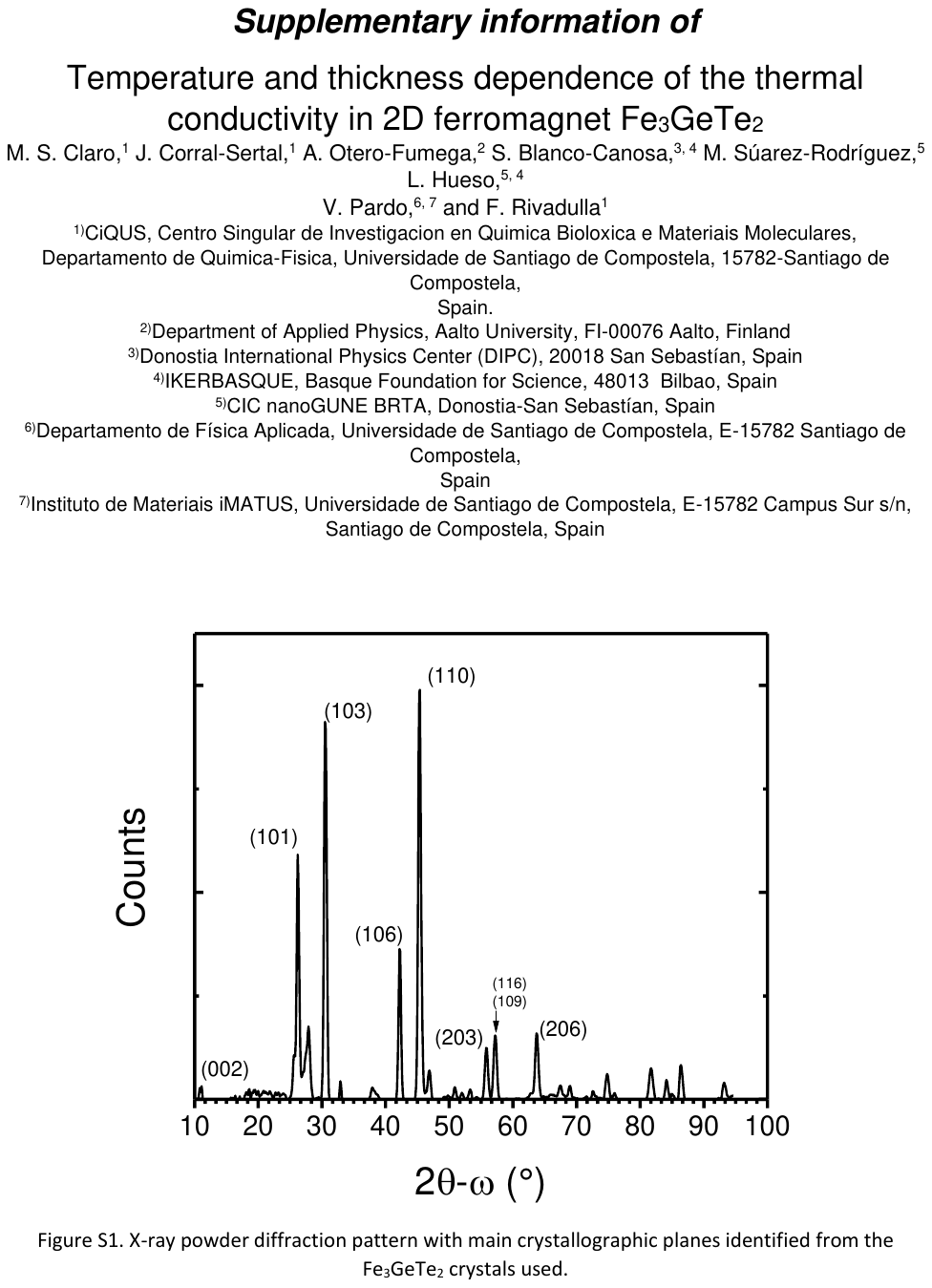}}

\begin{document}


\begin{abstract}
The emergence of symmetry-breaking orders such as ferromagnetism and the weak interlayer bonding in van der Waals materials, offers a unique platform to engineer novel heterostructures and tune transport properties like thermal conductivity. Here, we report the experimental and theoretical study of the cross-plane thermal conductivity, $\kappa_\perp$, of  the van der Waals 2D ferromagnet Fe$_3$GeTe$_2$.  
We observe a non-monotonic increase of $\kappa_\perp$ with the thickness and a large suppression in artificially-stacked layers, indicating a diffusive transport regime with ballistic contributions. These results are supported by the theoretical analyses of the accumulated thermal conductivity, which show an important contribution of phonons with mean free paths between 10 and 200 nm.
Moreover, our experiments show a reduction of the $\kappa_\perp$ in the low-temperature ferromagnetic phase occurring at the magnetic transition. 
The calculations show that this reduction in $\kappa_\perp$ is associated with a decrease in the group velocities of the acoustic phonons and an increase in the phonon-phonon scattering of the Raman modes that couple to the magnetic phase. 
These results demonstrate the potential of van der Waals ferromagnets for thermal transport engineering. 

\end{abstract}



The electric-field control of the  conductivity of atomic-thick graphene\cite{Novoselov2004,Novoselov2005Nature}, shortly afterwards extended to NbSe$_2$ and MoS$_2$\cite{NovoselovPNAS}, opened up new possibilities for materials properties manipulation in the novel world of 2D van der Waals (vdW) materials and heterostructures\cite{novoselov_2d_2016}. 
Particularly, on vdW materials, the extreme bonding anisotropy is  translated into a giant anisotropy also in the thermal transport, where the in-plane thermal conductivity $\kappa_\parallel$ is much larger than the cross-plane one  $\kappa_\perp$,
\cite{PhysRevMaterials.2.064005} despite the prediction of phonon mean-free paths (mfp) of the order of several tens of  nanometers across the weakly bonded planes
\cite{sood_quasi-ballistic_2019,vakulov_ballistic_2020}. Defects and imperfect layer  stacking result in a mixed contribution of ballistic transport (large mfp, coherent phonons) and diffusive transport (small mfp), which reduces very much the thermal conductivity across the 2D  planes.\cite{Kim2021,sood_quasi-ballistic_2019}


A particularly interesting 2D material regarding heat dissipation is the itinerant ferromagnet Fe$_3$GeTe$_2$ (FGT): charge doping through Li$^+$-intercalation modulates its magnetic anisotropy and increases T$_C$ up to room temperature,\cite{Ding_Nature} while a strong spin-phonon coupling \cite{Raman_AdvFuncMat} produces a significant effect of magnetic ordering on the thermal conductivity, opening the door to gate-tunable 2D thermal devices. First-principles calculations in other 2D magnetic materials, like  2H-VSe$_2$, CrI$_3$, FeX$_3$ and RuX$_3$ (X=Cl, Br, I)\cite{wu_giant_2023,qin_giant_2020,zhao_heat_2016,liu_effects_2021}, have predicted a large change of the thermal conductivity in their magnetically ordered phase as well, although an experimental confirmation of such a large switching of the thermal conductivity associated to magnetic ordering in 2D materials is lacking.

In this work, we report experimental measurements combined with a theoretical analysis of the thickness and temperature dependence of the thermal conductivity in FGT. We have observed a non-monotonic increase of the cross-plane thermal conductivity with thickness, characteristic of a mixed ballistic propagation of long mfp phonons with diffusive transport, as well as a large drop of the thermal conductivity in the magnetically ordered phase. Both effects can be understood by our \emph{ab initio} analysis of the thermal conductivity based on density functional theory (DFT) calculations. We also show that artificial layer stacking is a very effective way of blocking the cross-plane thermal transport in FGT. 

\section{Results and Discussion}

\begin{figure}[h!]
    \centering
    \includegraphics[width=\linewidth]{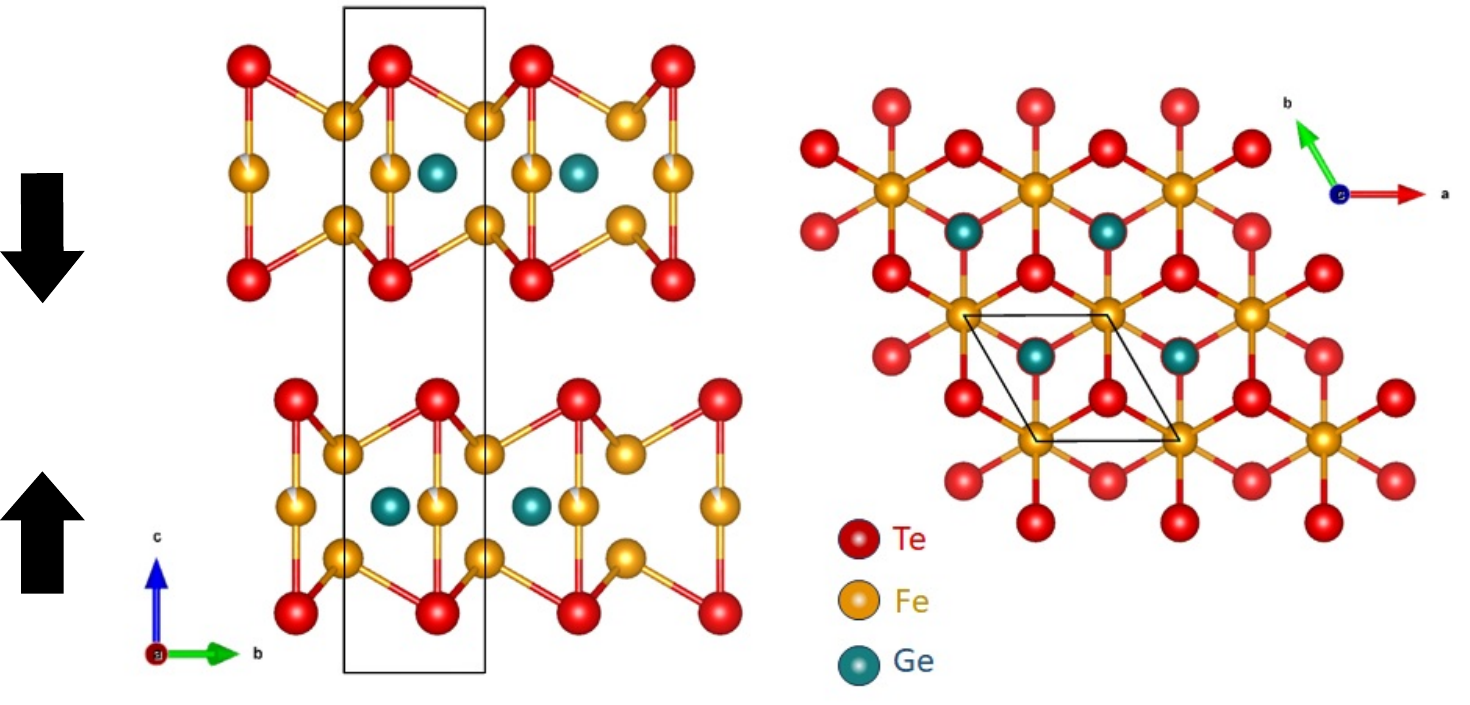}
    \caption{Crystal structure of $Fe_3GeTe_2$: a) lateral view of the structure. The cell is formed by two layers, shifted with respect to each other. In the low temperature magnetic phase the layers become ferromagnetic (FM) with an antiferromagnetic interlayer coupling. b) Top view of the structure showing the hexagonal symmetry of the ab plane. Fe, Ge and Te atoms are shown in gold, purple and green, respectively. 
    }
    \label{Crystal}
\end{figure}

FGT is a 2D itinerant ferromagnet with T$_C$$\approx$ 200K, which decreases with the number of layers, but retains the magnetic order down to the single-layer limit.\cite{fei_two-dimensional_2018} 
Neutron diffraction data support a ferromagnetic (FM) order also along the c-axis \cite{May2016}, although theoretical calculations and analysis of experimental magnetic susceptibility, suggested an antiferromagnetic (AF) stacking below T$_C$$\approx$ 152 K \cite{Yi_2017}.
From the structural point of view, the material is weakly bonded off the plane via van der Waals interactions, which facilitates its mechanical exfoliation and transfer of few-layer thick flakes to a  substrate. The unit cell consists of two vdW planes with hexagonal symmetry, each formed by 3 Fe atomic planes (see Fig. \ref{Crystal}).

\begin{figure}[h!]
    \centering
    \includegraphics[width=\linewidth]{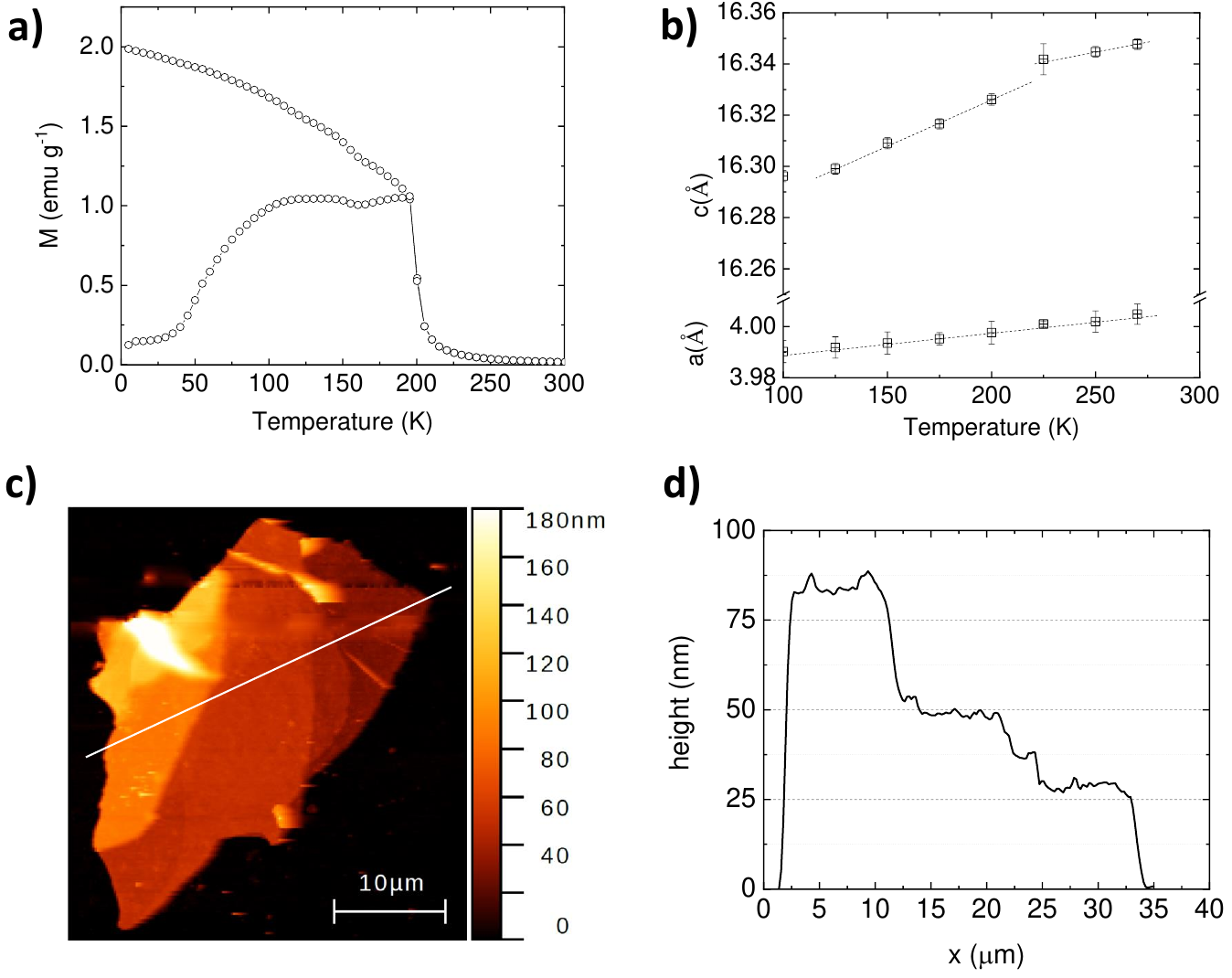}
    \caption{Temperature dependence of the magnetization, a), and lattice parameters, b), measured for a bulk crystal of FGT. c) Atomic force microscopy (AFM) image of one flake transferred to the surface of a (0001) sapphire substrate. The corresponding height profile along the line in c) is shown in panel d).  }
    \label{FigCharac}
\end{figure}

The crystals for this study were obtained from hq-graphene (hqgraphene.com; see the X-ray powder diffraction data in Figure S1 in the supplementary information). DC magnetization data of bulk crystals show that T$_C$$\approx$200 K, as expected for fully stoichiometric crystals (Figure \ref{FigCharac}a). Temperature-dependent X-ray analysis shows a change in the slope of the c-axis parameter at T$_C$ (Figure \ref{FigCharac}b), but no change in the space group of the crystal.

Few-layer thick flakes of FGT were prepared by mechanical exfoliation and transferred to (0001) sapphire substrates, using PDMS stamping\cite{spirito_tailoring_2022}. Transferred FGT flakes have tens of microns in lateral dimensions and thicknesses ranging from 15 nm to 250 nm (Figure \ref{FigCharac} c,d). The flakes are always thicker than 5 layers, considered to be the border between 2D and 3D magnetism,\cite{Tan2018} so that the comparison with the bulk calculations is justified.
Since vdW materials present high anisotropy between the conductivity in-plane $\kappa_\parallel$ ($\kappa_{xx}$,  $\kappa_{yy}$) and cross-plane $\kappa_\perp$ ($\kappa_{zz}$), their values will be considered separately.

Thermal conductivity was measured by Frequency Domain Thermoreflectance (FDTR), using a $\approx$60 nm thick layer of Au as a transducer\cite{yang_thermal_2013}. To extract $\kappa$ and the TBC from the FDTR phase-shift curves we fitted the most common model wherein total energy conservation and energy transfer between layers are imposed by a transfer matrix, as explained elsewhere\cite{schmidt_frequency-domain_2009}. To reduce the number of fitting parameters, the thickness of the Au layer was measured by X-ray reflectivity, and its thermal conductivity was estimated from the sheet electrical resistance measured by van der Paaw method, and the Wiedemann Franz law. The thickness of the FGT flakes was measured by atomic force microscopy (AFM). Heat capacities were taken from the literature and kept fixed for each temperature in all fittings.\cite{Liu2019} The thermal conductivity of the substrate was measured and confirmed with the values from the literature.\cite{LangenbergAPLMAT} In this way, the free parameters in the fittings are reduced to $\kappa_\perp$ of FGT (we observed that the sensitivity to $\kappa_\parallel$, is very low, and it has a negligible influence in $\kappa_\perp$), and the TBC  between Au/FGT, G1, and between FGT/Al$_2$O$_3$, G2. We considered the initial values for G1 $\approx$ 30-40~MWm$^{-2}$K$^{-1}$, similar to Au/MoS$_2$\cite{sood_quasi-ballistic_2019}, and G2 $\approx$ 25$~$MWm$^{-2}$K$^{-1}$, as reported for MoS$_2$/Al$_2$O$_3$  \cite{zheng_nonequilibrium_2022}, and MoS$_2$/SiO$_2$\cite{sood_quasi-ballistic_2019} interfaces.

\begin{figure}[h!]
    \centering
    \includegraphics[width=\linewidth]{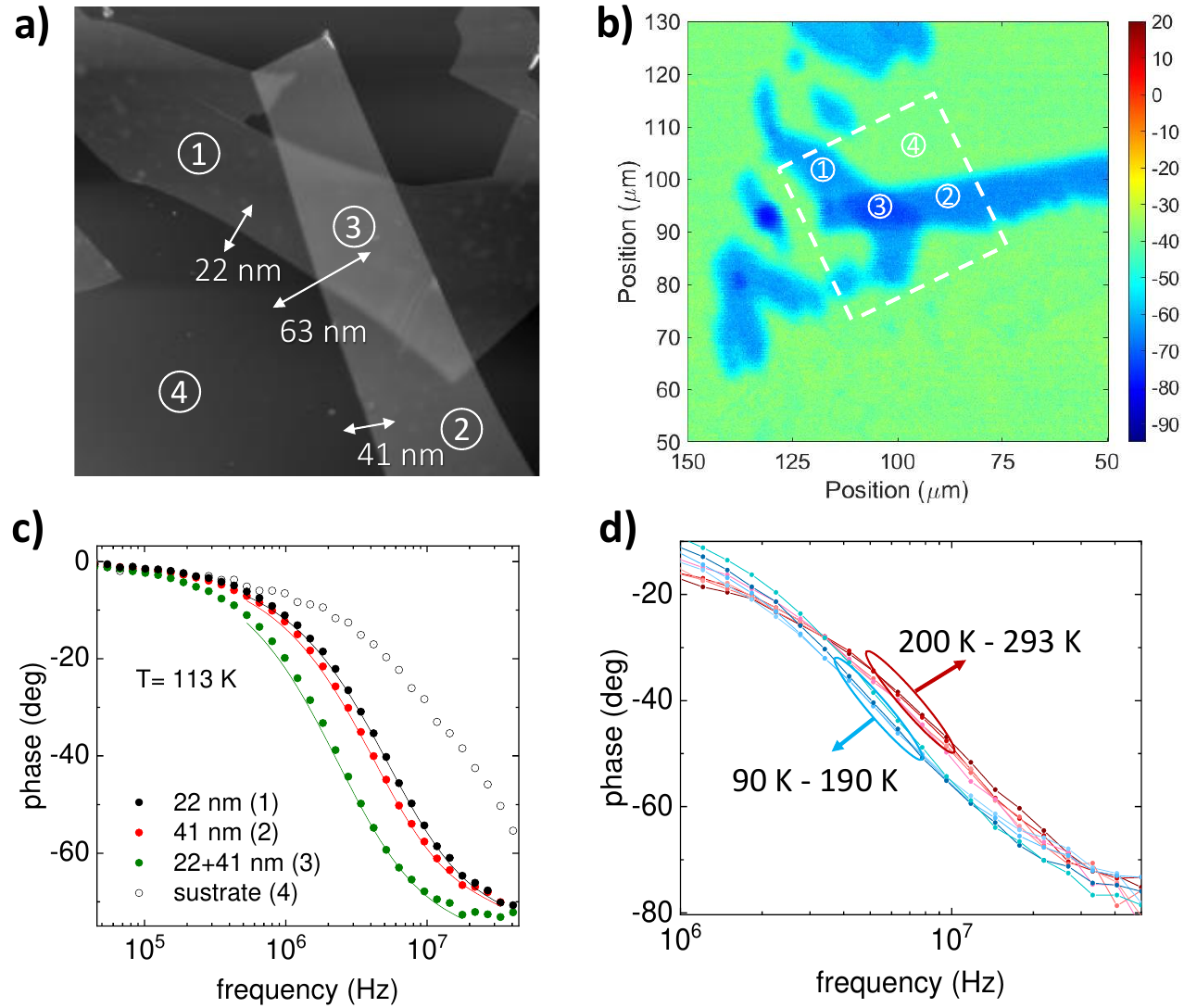}
        \caption{a) 30  $\times$ 30 $\mu m$  AFM topography  of two partially overlapping flakes . b) Phase-shift map at 20 MHz of the same region (enclosed within the square) observed in a), with the corresponding points marked. In this image, the flakes are already covered with 60 nm of Au for FDTR measurements. c) Phase-shift vs frequency curves for the three points marked in a) and b), along with the fitting to the thermal model. The curve of the substrate, as a reference, is also shown. d) Phase-shift vs frequency curves of point 2 at different temperatures. There is a large change around 200 K, associated to the magnetic ordering temperature (see text).}
    \label{FigAFMPhase}
\end{figure}

 \begin{figure}[h!]
    \centering
    \includegraphics[width=\linewidth]{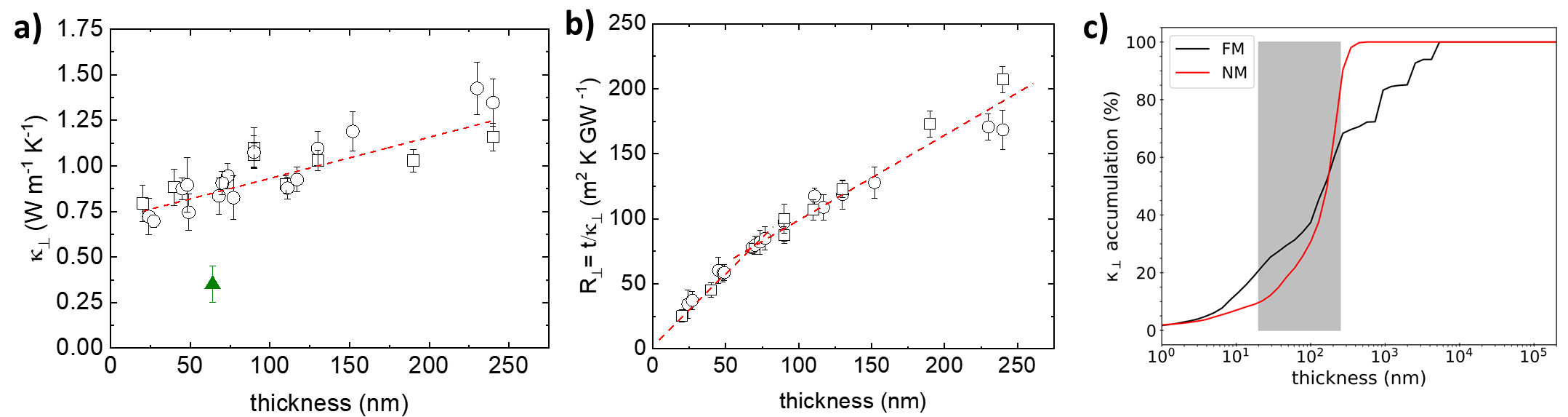}
    \caption{Measured thermal conductivity a), and thermal resistance R$_\perp$=t/$\kappa_\perp$, b) at room temperature for different flakes with varying thicknesses. Circles and squares correspond to different sets of crystals, transferred to different substrates. The green solid triangle in a) corresponds to $\kappa_\perp$ at point 3 in Fig. \ref{FigAFMPhase}a), the region of superposition of the two flakes. The dotted lines are linear fittings. c) Accumulated $\kappa_\perp$ as a function of the phonon mean free path, at 300 K. The shaded area shows the thickness range of the flakes studied by FDTR in this work.}
    \label{FigResThickness}
\end{figure}

On the other hand, the variability of G2 between mechanically transferred flakes may be an important source of error. For that reason, multilayer flakes like those shown in the Fig. \ref{FigCharac} c) are important to reduce problems associated with the variability of G2, as they allow the measurement of $\kappa_\perp$ for different thicknesses, with the same FGT/sapphire interface.

Initial values of $\kappa_\perp$ of 1$~$ Wm$^{-1}$K$^{-1}$, typical for other 2D materials, were used for an estimation of the sensitivity to different parameters in different frequency ranges. 
The spot size was varied between a 1/$e^2$ diameter $\approx$4$\mathrm{\mu}$m and 11$\mathrm{\mu}$m, for achieving better sensitivity to TBC and $\kappa_\perp$.
The fittings shown in Fig. \ref{FigAFMPhase}c) to obtain the $\kappa_\perp$ and TBCs reported in this work were performed from 1 MHz to 50 MHz, where the sensitivity for these parameters is maximum (see Figure S2 on Supplementary Information). 

Figure \ref{FigAFMPhase} a) shows the AFM topography of two partially overlapping flakes of thickness 22 nm and 41 nm each. The 100 $\times$ 100 $\mu m$ phase shift map at 20 MHz shows the variations in the contrast due to the differences in $\kappa_\perp$ and TBC. The whole frequency phase-shift spectra for each point marked in a) and b) are presented in Figure \ref{FigAFMPhase} c-d) at different temperatures, demonstrating good sensitivity to thickness and temperature.

 The thickness dependence of $\kappa_\perp$ at room temperature is shown in Figure \ref{FigResThickness}a). There is an increase of $\kappa_\perp$ with the thickness of the sample, of the order of $\approx$0.5 W/m K in a range of $\approx$200 nm. Although small, this is of the same order of magnitude as that reported for other van der Waals materials, like MoS$_2$\cite{sood_quasi-ballistic_2019} or SnSe$_2$,\cite{XiaoSnSe2}and it is consistent with our DFT calculations; Figure \ref{FigResThickness}c). The calculated  accumulated $\kappa_{\perp}$, shows that more than 50\% of the heat at 300 K is carried by phonons with a mean-free path larger than $\approx$200 nm, suggesting an important contribution from ballistic phonons along the c-axis, as in other vdW structures\cite{sood_quasi-ballistic_2019,li20202d}.

In the case of pure ballistic transport, phonons can propagate without thermal resistance inside the material, so that R$_\perp$=R$_{int}$+t/$\kappa_\perp$ should be a constant, independent of thickness.
However, the measured experimental cross-plane thermal resistance, R$_\perp$, also increases with the thickness (Figure \ref{FigResThickness}b). On the other hand, in a purely diffusive regime, R$_\perp$(t) is linear with constant slope =1/$\kappa_\perp$.\cite{Kim2021} For FGT, R$_\perp$ increases linearly with thickness above $\approx$60 nm, giving $\kappa_\perp$$\approx$1.9(1) Wm$^{-1}$K and R$_{int}$$\approx$46 m$^2$K/GW, but it deviates from this behavior for thinner samples, with a vanishing resistance as t$\rightarrow$0. The change in slope suggests some thickness-dependent contribution, and although the data in Figure \ref{FigResThickness}b) seem to extrapolate to zero, the thinner samples measured in this work are t$\approx$25 nm, so we cannot exclude a small finite value of R$_\perp$ close to the monolayer limit (note that a residual value as small as $\approx$10 m$^2$K/GW has been reported for a few monolayers of MoS$_2$).\cite{sood_quasi-ballistic_2019}

We also measured $\kappa_\perp$ in the superposition region of two crystals, point 3 in Figure \ref{FigAFMPhase}a): $\kappa_\perp$ is substantially reduced in the overlapping region of total thickness 63 nm (green triangle in Figure \ref{FigResThickness}a). Actually, the phase-shift curve of point 3 can be fitted with two layers, of 22 and 41 nm each, with their corresponding $\kappa_\perp$, and a very high interlayer thermal resistance between both flakes of $\approx$ 180 m$^2$K/GW (TBC$\approx$ 10-12 MW m$^{-2}$K$^{-1}$). This supports the contribution of phonons with a mfp larger than the thickness of the individual layers, evidencing the mixed contribution of diffusive and ballistic phonons to $\kappa_\perp$ in FGT.

It is also remarkable that the value of the TBC between the two FGT flakes is of the same order of magnitude as reported for interfaces between dissimilar 2D materials, like graphene/MoS$_2$ or MoS$_2$/WSe$_2$,\cite{vaziri_ultrahigh_2019} although in this case, the large interfacial resistance occurs between films of the same composition, without any mass-density or compositional mismatch. We have neither observed any dependence of TBC nor $\kappa_\perp$ on the relative orientation of the superposed FGT flakes.

\begin{figure}[h]
    \centering
    \includegraphics[width=\linewidth]{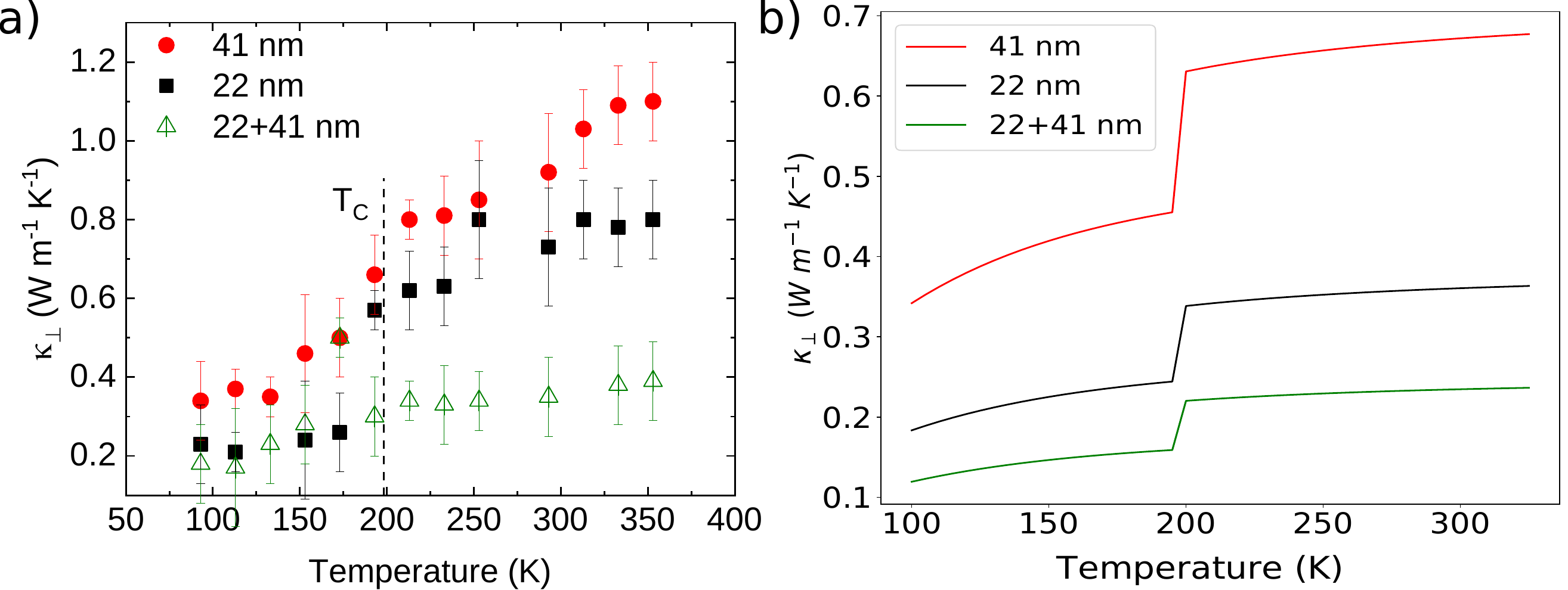}
    \caption{ Experimental a) and theoretical b) temperature dependence of the thermal conductivity of two flakes with thicknesses 22 nm, 41 nm and the superposition of both 22+41 nm, corresponding to points 1, 2 and 3 in Fig. \ref{FigAFMPhase}a) respectively. 
    }
    \label{FiResMagPhase}
\end{figure}

\begin{figure}[t]
    \centering
    \includegraphics[width =0.9\linewidth]{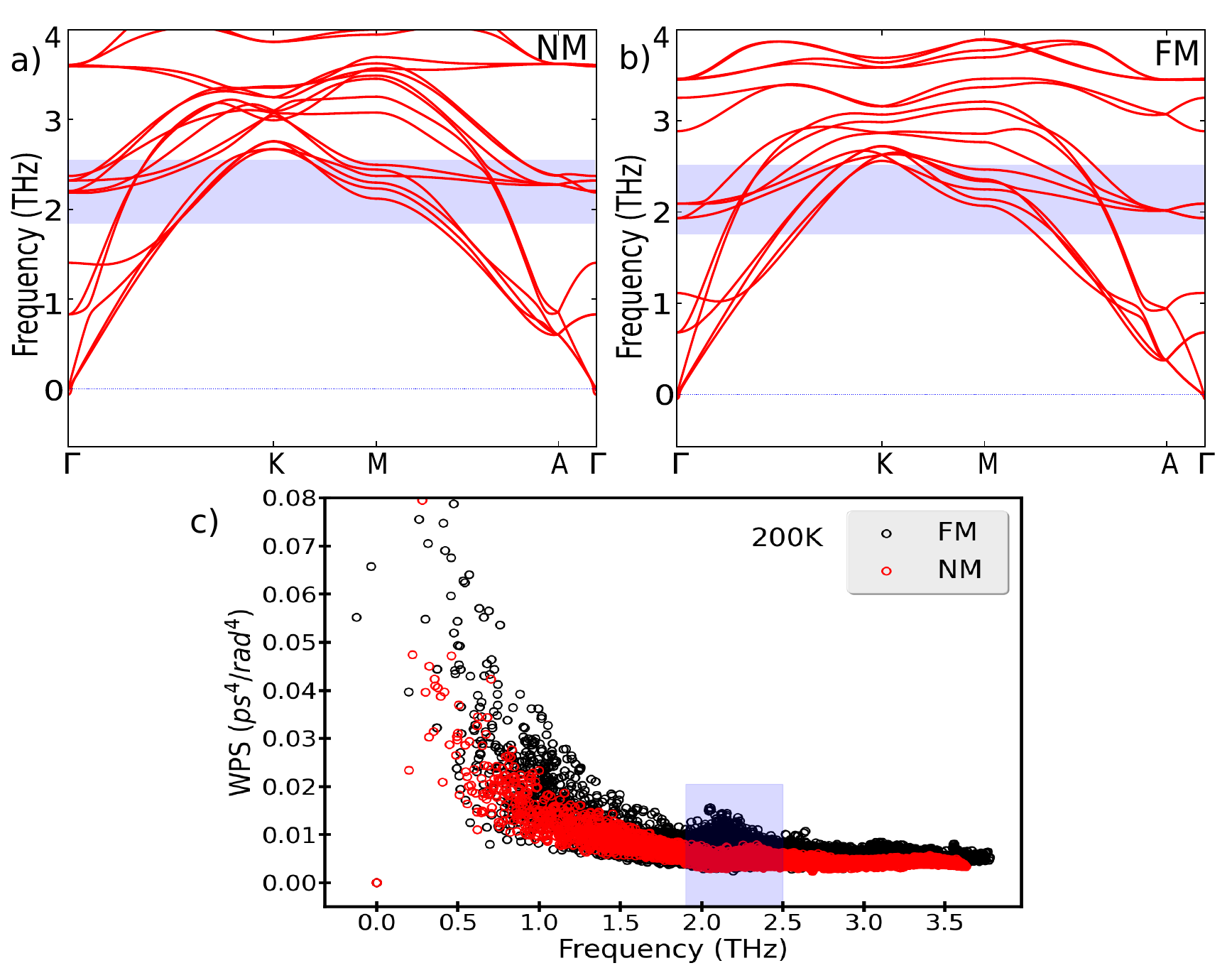}
    \caption{ Phonon band diagrams for the magnetic (FM) a) and non magnetic (NM)  b) ordering, and weighted phase  space (WPS) available for three-phonon processes as a function of the frequency c). The shaded area in c) corresponds with the highlighted region in both band diagrams. The FM ordering shows an enhanced peak compared to the NM ordering in the WPS around 2.1 THz associated to Raman modes, leading to more scattering processes causing a reduction in the cross-plane thermal conductivity. } 
    \label{fig:WP3}
\end{figure}

Finally, the experimental temperature dependence of $\kappa_\perp$  is shown in  Fig. \ref{FiResMagPhase}a), for two different thicknesses (points 1 and 2 in Fig. \ref{FigAFMPhase}a).  A reduction of $\kappa_\perp$ between 25$\%$ and 65$\%$ occurs below $T_C$ in the transition to the magnetic phase. Note that the jump in $\kappa_\perp$  is clearly observed in the raw phase-shift curves (Fig. \ref{FigAFMPhase}d) and therefore it cannot be attributed to fitting artifacts.
The temperature dependence of $\kappa_\perp$ in the region of superposition of the two flakes (point 3 in Figure \ref{FigAFMPhase}a) is also shown. In this case, $\kappa_\perp$ remains very small, with a weak temperature dependence in the whole interval, without any substantial change around T$_C$, suggesting that in this case the flakes act as independent resistances.

It is common in magnetic and ferroelectric materials that the formation of domain walls causes a reduction of thermal conductivity due to phonon scattering on domain boundaries\cite{huang_influence_2015,nakayama_above-room-temperature_2021}. However, the $\kappa_\perp$ obtained is robust to external magnetic fields up to 50 mT, applied with a strong toroidal permanent magnet (see Suppl. Information Figure S4 and S5). Based on previous reports\cite{birch_history-dependent_2022} this applied field should be enough to switch between stripe domains and uniformly magnetized states; the negligible effect of the magnetic field on $\kappa_\perp$ indicates that the in-plane magnetic domains are not the cause of the sudden change of $\kappa_\perp$ at $T_c$.  
We have also discarded as a cause of the jump on $\kappa_\perp$ the eventual changes in the crystal structure, since the powder X-Ray diffraction of the original bulk crystal revealed only a small change in the c-axis lattice parameter and thermal expansion, without any crystallographic transformation (Figure \ref{FigCharac}b).

In order to shed light on the observed experimental behavior, we have computed 
the temperature dependence of $\kappa_{\perp}$ in the small-grain limit for the different experimental cases (Fig. \ref{FiResMagPhase}b). The thickness of the flakes was used as the boundary length for the 41 nm and 22 nm flakes. For the superposition region, 22+41 nm, $\kappa_\perp^{22+41nm}$ was estimated considering that the two flakes of 22 and 41 nm act as two independent resistances in series $\kappa_\perp^{22+41nm}=\kappa_\perp^{22nm}\kappa_\perp^{41nm}/(\kappa_\perp^{22nm}+\kappa_\perp^{41nm})$. For temperatures above (below) $T_C$ the non-magnetic (NM) (ferromagnetic (FM) layers ordered antiferromagnetically) solution was considered when estimating the thermal conductivity.  
Our calculations show a $\approx$30 \% drop in the thermal conductivity between the NM and FM phases in good agreement with the experimental observations.
Note that the theoretical underestimation of the thermal conductivity is related to the limitations of the small-grain limit used in the calculations, in which the boundary scattering is overestimated, especially at higher temperatures and for larger samples. However, all the qualitative features are well captured on it. 

To understand the reduction of $\kappa_\perp$ when entering the magnetic phase we can analyze the phonon band structures and the weighted phase space (WPS) available for three-phonon processes for the non-magnetic and the ferromagnetic configurations (Fig. \ref{fig:WP3}). 
This last quantity gives an idea of the frequencies involved in phonon scattering processes that are different in the FM and NM states. 
From the phonon band structures, we can observe that the acoustic phonons undergo a shift toward lower frequencies, especially in the A-$\Gamma$ path. This is related to a decrease in group velocities in the out-of-plane direction and hence in the thermal conductivity\cite{PhysRevMaterials.4.033606}. 

Moreover, the FM ordering shows an increase in the WPS, specifically a peak around 2 THz that is substantially different from the NM calculation.
This peak is related to the phonon modes highlighted Fig. \ref{fig:WP3} a) and b) and corresponds to two E$_{1g}$ Raman active modes. In the magnetic phase, these modes are about 0.3 THz lower in energy, showing more crossings with the acoustic modes. This results in additional scattering processes and causes a reduction in the thermal conductivity. The large coupling between Raman-active modes and a particular magnetic order has been reported in other two-dimensional magnets.\cite{huang2020tuning} Here, we observe that this results, together with the decrease in the group velocities of the acoustic modes, in a reduction of the lattice thermal conductivity in the magnetic phase.



\section{Conclusion}

To summarize, we have combined experimental FDTR and ab initio calculations to demonstrate that the cross-plane thermal conductivity of 2D ferromagnet FGT presents a mixed contribution of diffusive and ballistic phonons. We have also shown that $\kappa_\perp$ presents an abrupt reduction below the Curie temperature, due to additional phonon scattering produced by a downshift in the frequency of acoustic and Raman-active optical phonons in the magnetic phase.
Finally, artificial stacking of a few-layer thick FGT is a very useful way of reducing the cross-plane thermal conductivity in this material.

\section{Experimental and Computational Details}

Thermal conductivity was measured by Frequency Domain Thermoreflectance (FDTR)\cite{yang_thermal_2013}, using a 435 nm pump laser (1 mW) and a 532 nm probe laser (3 mW), respectively. A 60 nm gold thin film deposited by sputtering works as a reflective transducer. The fitting model considers Fourier heat conduction: the heat flux $q=-\kappa \nabla T$ where $\kappa$ is a tensor to account for the material thermal conductivity anisotropy. The sample temperature is controlled inside a cold-finger optical cryostat, down to 80 K. The whole stage is mounted on a piezoelectric table which allows $\mathrm{\mu m}$ precision location of the laser spots on the sample. To promote the adhesion of FGT to the sapphire substrate, the samples were annealed under vacuum at 100ºC before the experiments.

$\kappa_\perp$ in the magnetic and non-magnetic phases of bulk FGT are calculated within a DFT\cite{hohenberg1964inhomogeneous,kohn1965self} framework using the  {\sc VASP} code.  \cite{kresse1993ab, kresse1996efficiency, kresse1996efficient} For all calculations, we have performed a full relaxation of the structure (both atomic positions and lattice parameters were optimized) with a mesh of 16 $\times$ 16 $\times$ 3 k-points in the irreducible wedge of the Brillouin zone. The exchange-correlation potential chosen was the generalized gradient approximation in the Perdew-Burke-Ernzerhof scheme. \cite{gga_pbe} The second order interatomic force constants (IFCs) were determined using the Phonopy code  \cite{phonopy, phonopy-phono3py-JPSJ} in a  2$\times$2$\times$2 supercell with a k-mesh of 8$\times$8$\times$2 with no further relaxation of cell shape or volume. Third order anharmonic IFCs were computed using the machinery of the ShengBTE code,  \cite{ShengBTE_2014} considering interactions to third neighbors in a 2$\times$2$\times$2 supercell. The lattice thermal conductivity was calculated by solving the Boltzmann Transport Equation (BTE) for phonons by an iterative self-consistent  method implemented in the ShengBTE code within a mesh of 36$\times$36$\times$8 $q$-points and a scalebroad parameter of 0.1.

\begin{acknowledgement}
This work has received financial support from Ministerio de Economía y Competitividad (Spain), projects PID2019-104150RB-I00 and PID2021-122609NB-C22, Xunta de Galicia (Centro singular de investigación de Galicia accreditation 2019-2022, ED431G 2019/03), the European Union (European Regional Development Fund-ERDF). We thank the CESGA (Centro de Supercomputaci\'on de Galicia for the computational facilities. A.O.F. acknowledges the Academy of Finland Project No. 349696.
\end{acknowledgement}

\begin{suppinfo}

Additional figures with X-ray powder diffraction pattern of exfoliated crystal, FTDR sensitivity analysis and details of FDTR measurements with magnetic field.

\end{suppinfo}

\bibliography{references}

\end{document}